\documentclass[aip,rsi]{revtex4-1}
\usepackage{hyperref} 
\usepackage{graphicx}
\usepackage{xcolor}
\usepackage{textcomp}           % for non-italic mu

\newcommand{\microrad}{\textmu{}rad }
\newcommand{\microradnospace}{\textmu{}rad}

\begin{document}

\title{Characterization of thermal effects in the Enhanced LIGO Input Optics}

\author{Katherine L. Dooley}
\email[Author to whom correspondence should be addressed. Electronic
mail: ]{kate.dooley@aei.mpg.de}
\altaffiliation[Current address: ]{Albert-Einstein-Institut, Max-Planck-Institut f\"{u}r Gravitationsphysik, D-30167
Hannover, Germany}
\affiliation{University of Florida, Gainesville, FL 32611, USA}

\author{Muzammil A. Arain}
\altaffiliation[Current address: ]{KLA-Tencor, Milpitas, California 95035, USA}
\affiliation{University of Florida, Gainesville, FL
32611, USA}

\author{David Feldbaum}
\affiliation{University of Florida, Gainesville, FL
32611, USA}

\author{Valery V. Frolov}
\affiliation{LIGO - Livingston Observatory, Livingston, LA 70754, USA}

\author{Matthew Heintze}
\affiliation{University of Florida, Gainesville, FL
32611, USA}

\author{Daniel Hoak}
\altaffiliation[Current address: ]{University of Massachusetts--Amherst, Amherst, MA 01003, USA}
\affiliation{LIGO - Livingston Observatory, Livingston, LA 70754, USA}

\author{Efim A. Khazanov}
\affiliation{Institute of Applied Physics, Nizhny Novgorod 603950,
 Russia}

\author{Antonio Lucianetti}
\altaffiliation[Current address: ]{\'{E}cole Polytechnique, 91128 Palaiseau Cedex,
  France}
\affiliation{University of Florida, Gainesville, FL
32611, USA}

\author{Rodica M. Martin}
\affiliation{University of Florida, Gainesville, FL
32611, USA}

\author{Guido Mueller}
\affiliation{University of Florida, Gainesville, FL
32611, USA}

\author{Oleg Palashov}
\affiliation{Institute of Applied Physics, Nizhny Novgorod 603950,
 Russia}

\author{Volker Quetschke}
\altaffiliation[Current address: ]{The University of Texas at
  Brownsville, Brownsville, TX 78520, USA}
\affiliation{University of Florida, Gainesville, FL
32611, USA}

\author{David H. Reitze}
\altaffiliation[Current address: ]{LIGO - California Institute of Technology, Pasadena, CA 91125, USA}
\affiliation{University of Florida, Gainesville, FL 32611, USA}

\author{R. L. Savage}
\affiliation{LIGO - Hanford Observatory, Richland, WA 99352, USA}

\author{D. B. Tanner}
\affiliation{University of Florida, Gainesville, FL
32611, USA}

\author{Luke F. Williams}
\affiliation{University of Florida, Gainesville, FL
32611, USA}

\author{Wan Wu}
\altaffiliation[Current address: ]{Nasa Langley Research Center, Hampton, VA 23666, USA}
\affiliation{University of Florida, Gainesville, FL
32611, USA}

\begin{abstract}
  We present the design and performance of the LIGO Input Optics
  subsystem as implemented for the sixth science run of the LIGO
  interferometers. The Initial LIGO Input Optics experienced thermal
  side effects when operating with 7~W input power. We designed,
  built, and implemented improved versions of the Input Optics for
  Enhanced LIGO, an incremental upgrade to the Initial LIGO
  interferometers, designed to run with 30~W input power. At four
  times the power of Initial LIGO, the Enhanced LIGO Input Optics
  demonstrated improved performance including better optical
  isolation, less thermal drift, minimal thermal lensing and higher
  optical efficiency. The success of the Input Optics design fosters
  confidence for its ability to perform well in Advanced LIGO.
\end{abstract}

%\pacs{xxx}

\maketitle

\section{Introduction}
The field of ground-based gravitational-wave (GW) physics is rapidly
approaching a state with a high likelihood of detecting GWs for the
first time in the latter half of this decade. Such a detection will
not only validate part of Einstein's general theory of relativity, but
initiate an era of astrophysical observation of the universe through
GWs. Gravitational waves are dynamical strains in space-time, $h =
\Delta L/L$, that travel at the speed of light and are generated by
non-axisymmetric acceleration of mass. A first detection is expected
to witness an event such as a binary black hole/neutron star merger
\citep{Abadie2010Predictions}.

\begin{figure}
\begin{centering}
\includegraphics[width=1.0\columnwidth]{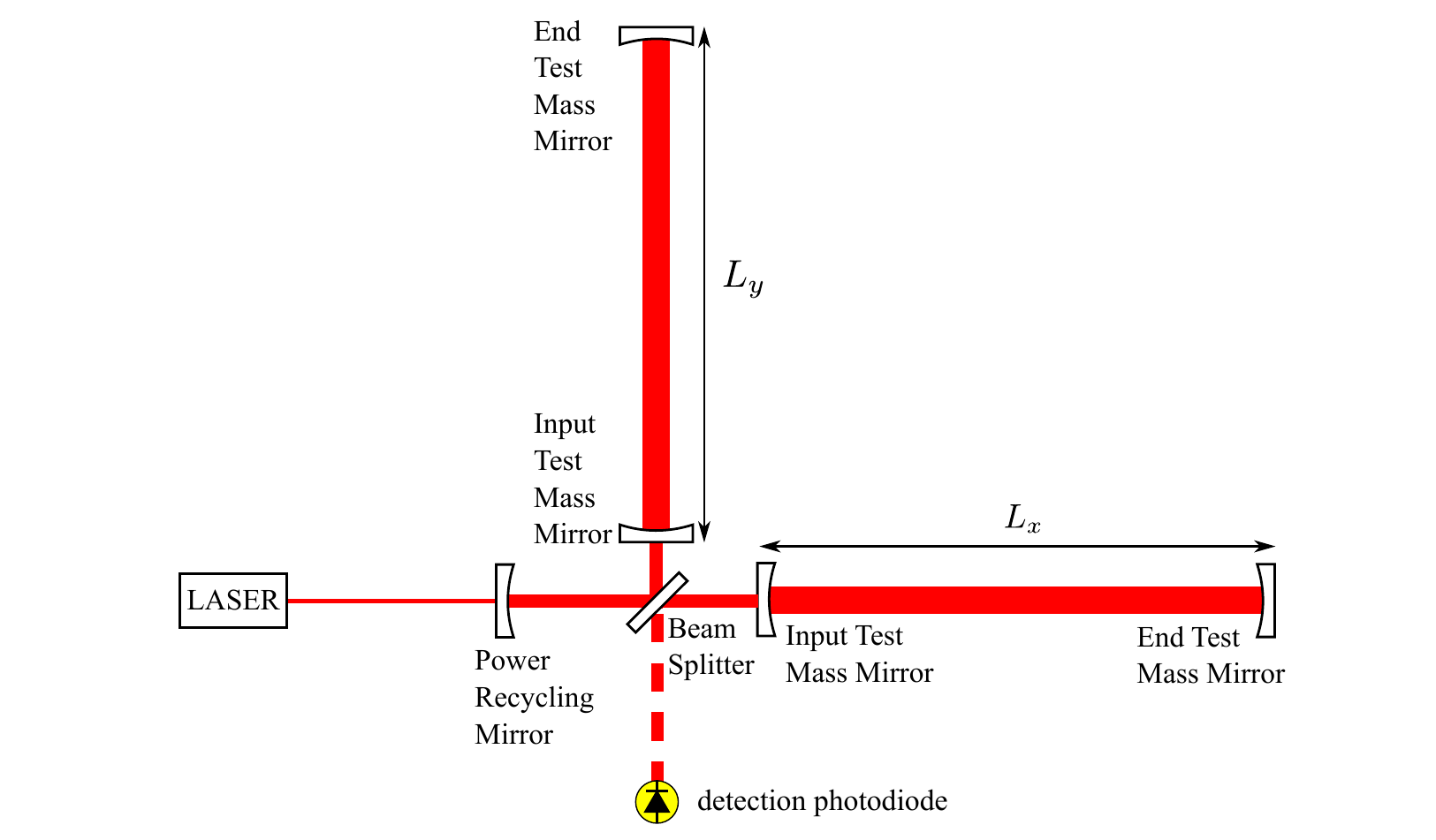}
\caption{(Color online) Optical layout of a Fabry-Perot Michelson
  laser interferometer, showing primary components. The four test
  masses, beam splitter and power recycling mirror are physically
  located in an ultrahigh vacuum system and are seismically
  isolated. A photodiode at the anti-symmetric port detects
  differential arm length changes.}
\label{fig:IFOschematic}
\end{centering}
\end{figure}

The typical detector configuration used by current generation
gravitational-wave observatories is a power-recycled Fabry-Perot
Michelson laser interferometer featuring suspended test masses in
vacuum as depicted in Figure \ref{fig:IFOschematic}. A diode-pumped,
power amplified, and intensity and frequency stabilized Nd:YAG laser
emits light at $\lambda = 1064$~nm. The laser is directed to a Michelson
interferometer whose two arm lengths are set to maintain destructive
interference of the recombined light at the anti-symmetric (AS)
port. An appropriately polarized gravitational wave will
differentially change the arm lengths, producing signal at the AS port
proportional to the GW strain and the input power. The Fabry-Perot
cavities in the Michelson arms and a power recycling mirror (RM) at
the symmetric port are two modifications to the Michelson
interferometer that increase the laser power in the arms and therefore
improve the detector's sensitivity to GWs.

A network of first generation kilometer scale laser interferometer
gravitational-wave detectors completed an integrated 2-year data
collection run in 2007, called Science Run 5 (S5). The instruments
were: the American Laser Interferometer Gravitational-wave
Observatories (LIGO)\citep{Abbott2009LIGO}, one in Livingston, LA with
4 km long arms and two in Hanford, WA with 4 km and 2 km long arms;
the 3 km French-Italian detector VIRGO\citep{Acernese2008Virgo} in
Cascina, Italy; and the 600~m German-British detector
GEO\citep{Luck2006Status} located near Hannover, Germany. Multiple
separated detectors increase detection confidence through signal
coincidence and improve source localization via waveform
reconstruction.

The first generation of LIGO, now known as Initial LIGO, achieved its
design goal of sensitivity to GWs in the 40--7000~Hz band, including a
record strain sensitivity of $2\times10^{-23}/\sqrt{\mathrm{Hz}}$ at
155~Hz. However, only nearby sources produce enough GW strain to
appear above the noise level of Initial LIGO and no gravitational wave
has yet been found in the S5 data. A second generation of LIGO
detectors, Advanced LIGO, has been designed to be at least an order of
magnitude more sensitive at several hundred Hz and above and to give
an impressive increase in bandwidth down to 10~Hz.  Advanced LIGO is
expected to open the field of GW astronomy through the detection of
many events per year \citep{Abadie2010Predictions}.  To test some of
Advanced LIGO's new technologies and to increase the chances of
detection through a more sensitive data taking run, an incremental
upgrade to the detectors was carried out after S5
\citep{Adhikari2006Enhanced}. This project, Enhanced LIGO, culminated
with the S6 science run from July 2009 to October 2010. Currently,
construction of Advanced LIGO is underway.  Simultaneously, VIRGO and
GEO are both undergoing their own upgrades \citep{Acernese2008Virgo,
  Luck2010Upgrade}.

The baseline Advanced LIGO design \citep{AdvLigoSysDesign} improves
upon Initial LIGO by incorporating improved seismic isolation
\citep{Robertson2004Seismic}, the addition of a signal recycling mirror
at the output port \citep{Meers1988Recycling}, homodyne readout, 
and an increase in available laser power from 8~W to 180~W. The substantial
increase in laser power improves the shot-noise-limited sensitivity,
but introduces a multitude of thermally induced side effects that must
be addressed for proper operation.

Enhanced LIGO tested portions of the Advanced LIGO designs so that
unforeseen difficulties could be addressed and so that a more
sensitive data taking run could take place. An output mode cleaner was
designed, built and installed, and DC readout of the GW signal was
implemented \citep{Fricke2011DC}. An Advanced LIGO active seismic
isolation table was also built, installed, and tested \cite[Chapter
5]{KisselThesis}. In addition, the 10~W Initial LIGO laser was
replaced with a 35~W laser \citep{Frede2007Fundamental}. Accompanying
the increase in laser power, the test mass Thermal Compensation System
\citep{Willems2009Thermal}, the Alignment Sensing and Control
\citep{DooleyAngular}, and the Input Optics were modified.

This paper reports on the design and performance of the LIGO Input
Optics (IO) subsystem in Enhanced LIGO, focusing specifically on its
operational capabilities as the laser power is increased to 30~W.
Substantial improvements in the IO power handling capabilities with
respect to Initial LIGO performance are seen.  The paper is organized
as follows.  First, in Section \ref{sec:role} we define the role of
the IO subsystem and detail the function of each of the
major IO subcomponents. Then, in Section \ref{sec:problems} we
describe thermal effects which impact the operation of the IO and
summarize the problems experienced with the IO in Initial
LIGO. In Section \ref{sec:design} we present the IO design for
Advanced LIGO in detail and describe how it addresses these
problems. Section \ref{sec:performance} presents the performance of
the prototype Advanced LIGO IO design as tested during Enhanced
LIGO. Finally, we extrapolate from these experiences in Section
\ref{sec:aLIGO} to discuss the expected IO performance in
Advanced LIGO. The paper concludes with a summary in Section
\ref{sec:summary}.

\section{Function of the Input Optics}
\label{sec:role}

\begin{figure*}
\begin{centering}
\includegraphics[width=1.0\textwidth]{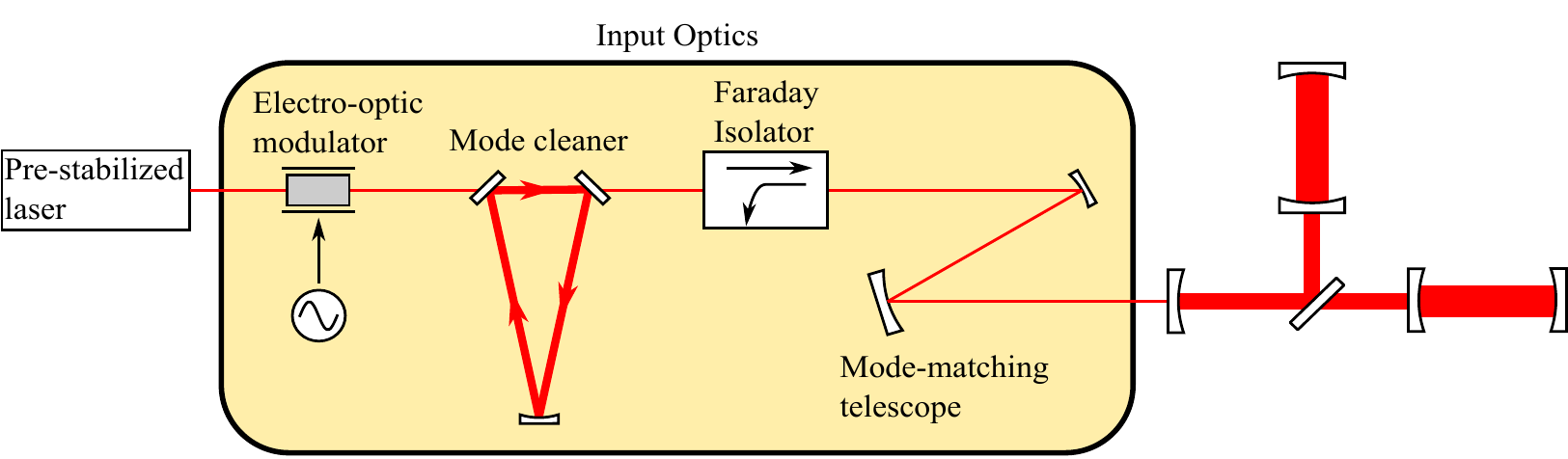}
\caption{(Color online) Block diagram of the Input Optics
  subsystem. The IO is located between the pre-stabilized laser and
  the recycling mirror and consists of four principle components: electro-optic
  modulator, mode cleaner, Farday isolator, and mode-matching
  telescope. The electro-optic modulator is the only IO component
  outside of the vacuum system. Diagram is not to scale.}
\label{fig:IOblock}
\end{centering}
\end{figure*}

The Input Optics is one of the primary subsystems of the LIGO
interferometers. Its purpose is to deliver an aligned, spatially pure,
mode-matched beam with phase-modulation sidebands to the
power-recycled Fabry-Perot Michelson interferometer. The IO also
prevents reflected or backscattered light from reaching the laser and
distributes the reflected field from the interferometer (designated
the \emph{reflected port}) to photodiodes for sensing and controlling
the length and alignment of the interferometer. In addition, the IO
provides an intermediate level of frequency stabilization and must
have high overall optical efficiency. It must perform these functions
without limiting the strain sensitivity of the LIGO interferometer.
Finally, it must operate robustly and continuously over years of
operation. The conceptual design is found in
Ref. \citep{Camp1996InputOutput}.

As shown in Fig. \ref{fig:IOblock}, the IO subsystem consists of four
principle components located between the pre-stabilized laser and the
power recycling mirror:
\begin{itemize}
\item electro-optic modulator (EOM)
\item mode cleaner cavity (MC)
\item Faraday isolator (FI)
\item mode-matching telescope (MMT)
\end{itemize}
Each element is a common building block of many optical experiments
and not unique to LIGO. However, their roles specific to the
successful operation of interferometry for gravitational-wave
detection are of interest and demand further attention. Here, we
briefly review the purpose of each of the IO components;
further details about the design requirements are in
Ref. \citep{Camp1997Input}.

\subsection{Electro-optic modulator} 
The Length Sensing and Control (LSC) and Angular Sensing and Control
(ASC) subsystems require phase modulation of the laser light at RF
frequencies. This modulation is produced by an EOM, generating
sidebands of the laser light which act as references against which
interferometer length and angle changes are measured
\citep{Fritschel2001Readout}. The sideband light must be either
resonant only in the recycling cavity or not resonant in the
interferometer at all. The sidebands must be offset from the carrier
by integer multiples of the MC free spectral range to pass through the
MC. 

\subsection{Mode cleaner}
Stably aligned cavities, limited non-mode-matched (junk) light, and a
frequency and amplitude stabilized laser are key features of any ultra
sensitive laser interferometer. The MC, at the heart of the
IO, plays a major role.

A three-mirror triangular ring cavity, the MC suppresses
laser output not in the fundamental TEM$_{00}$ mode, serving two major
purposes. It enables the robustness of the ASC because higher order
modes would otherwise contaminate the angular sensing signals of the
interferometer. Also, all non-TEM$_{00}$ light on the length sensing
photodiodes, including those used for the GW readout, contributes shot
noise but not signal and therefore diminishes the signal to noise
ratio. The MC is thus largely responsible for achieving an
aligned, minimally shot-noise-limited interferometer.

The MC also plays an active role in laser frequency
stabilization \citep{Fritschel2001Readout}, which is necessary for
ensuring that the signal at the anti-symmetric port is due to arm
length fluctuations rather than laser frequency fluctuations. In
addition, the MC passively suppresses beam jitter at frequencies above
10~Hz.

\subsection{Faraday isolator}
Faraday isolators are four-port optical devices which utilize the
Faraday effect to allow for non-reciprocal polarization switching of
laser beams.  Any backscatter or reflected light from the interferometer (due to
impedance mismatch, mode mismatch, non-resonant sidebands, or signal)
needs to be diverted to protect the laser from back propagating light,
which can introduce amplitude and phase noise.  This diversion of the
reflected light is also necessary for extracting length and angular
information about the interferometer's cavities. The FI fulfils
both needs.

\subsection{Mode-matching telescope}
The lowest-order MC and arm cavity spatial eigenmodes need to be
matched for maximal power buildup in the interferometer. The
mode-matching telescope is a set of three suspended concave mirrors
between the MC and interferometer that expand the beam from a radius
of 1.6 mm at the MC waist to a radius of 33 mm at the arm cavity
waist. The MMT should play a passive role by delivering properly
shaped light to the interferometer without introducing beam jitter or
any significant aberration that can reduce mode coupling.

\section{Thermal problems in Initial LIGO}
\label{sec:problems}
The Initial LIGO interferometers were equipped with a 10~W laser, yet
operated with only 7~W input power due to
power-related problems with other subsystems. The EOM was located in
the 10~W beam and the other components experienced anywhere up to 7~W
power. The 7~W operational limit was not due to the failure of the
IO; however, many aspects of the IO performance did degrade
with power.

\begin{figure*}
\begin{centering}
\includegraphics[width=1.0\textwidth]{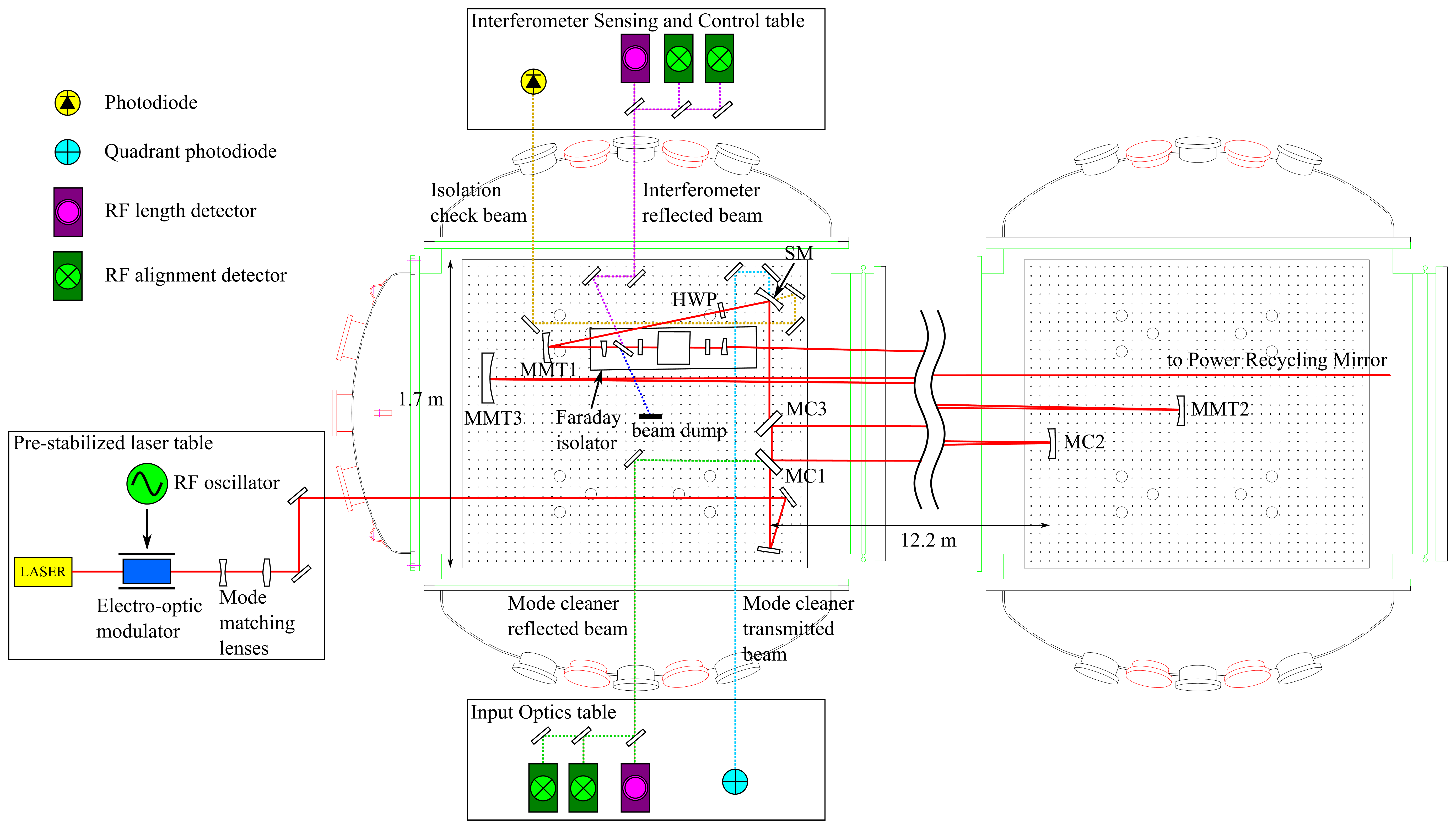}
\caption{(Color online) Enhanced LIGO Input Optics optical and sensing
  configuration. The HAM1 (horizontal access module) vacuum chamber is
  featured in the center, with locations of all major optics
  superimposed. HAM2 is shown on the right, with its components. These
  tables are separated by 12~m. The primary beam path, beginning at
  the pre-stabilized laser and going to the power recycling mirror, is
  shown in red as a solid line, and auxiliary beams are different
  colors and dotted. The MMTs, MCs, and steering mirror (SM) are
  suspended; all other optics are fixed to the seismically isolated
  table. The laser and sensing and diagnostic photodiodes are on
  in-air tables.}
\label{fig:IOschematic}
\end{centering}
\end{figure*}

One of the primary problems of the Initial LIGO IO
\citep{Adhikari1998Input} was thermal deflection of the back
propagating beam due to thermally-induced refractive index gradients
in the FI. A significant beam drift between the
interferometer's locked and unlocked states led to clipping of the
reflected beam on the photodiodes used for length and alignment
control (see Fig. \ref{fig:IOschematic}). Our measurements determined
a deflection of approximately 100~\microradnospace/W in the FI.  This
problem was
mitigated at the time by the design and implementation of an active
beam steering servo on the beam coming from the isolator.

There were also known limits to the power the IO could sustain.
Thermal lensing in the FI optics began to alter
significantly the beam mode at powers greater than 10~W, leading to a
several percent reduction in mode matching to the interferometer
\citep{UFLIGOGroup2006Upgrading}.  Additionally, absorptive FI
elements would create thermal birefringence, degrading the optical
efficiency and isolation ratio with power
\citep{Khazanov1999Investigation}.  The Initial LIGO New Focus
EOMs had an operational power limit of around
10~W. There was a high risk of damage to the crystals under the stress
of the 0.4~mm radius beam. Also, anisotropic thermal lensing with
focal lengths as severe as 3.3~m at 10~W made the EOMs unsuitable for
much higher power. Finally, the MC mirrors exhibited high
absorption (as much as 24 ppm per mirror)--enough that thermal lensing
of the MC optics at Enhanced LIGO powers would induce higher order
modal frequency degeneracy and result in a power-dependent mode
mismatch into the interferometer \citep{Bullington2008Modal,
  Arain2007Note}. In fact, as input power increased from 1~W to 7~W
the mode matching decreased from 90\% to 83\%.

In addition to the thermal limitations of the Initial LIGO IO, optical
efficiency in delivering light from the laser into the interferometer
was not optimal. Of the light entering the IO chain, only
60\% remained by the time it reached the power recycling
mirror. Moreover, because at best only 90\% of the light at the
recycling mirror was coupled into the arm cavity mode, room was left
for improvement in the implementation of the MMT.

\section{Enhanced LIGO Input Optics Design}
\label{sec:design}
The Enhanced LIGO IO design addressed the thermal effects that
compromised the performance of the Initial LIGO IO, and accommodated
up to four times the power of Initial LIGO. Also, the design was a
prototype for handling the 180~W laser planned for Advanced
LIGO. Because the adverse thermal properties of the Initial LIGO IO
(beam drift, birefringence, and lensing) are all attributable
primarily to absorption of laser light by the optical elements, the
primary design consideration was finding optics with lower absorption
\citep{UFLIGOGroup2006Upgrading}. Both the EOM and the FI were
replaced for Enhanced LIGO. Only minor changes were made to the MC and
MMT. A detailed layout of the Enhanced LIGO IO is shown in Figure
\ref{fig:IOschematic}.

\subsection{Electro-optic modulator design}
We replaced the commercially-made New Focus 4003 resonant phase
modulator of Initial LIGO with an in-house EOM design and
construction. Both a new crystal choice and architectural design
change allow for superior performance.

The Enhanced LIGO EOM design uses a crystal of rubidium titanyl
phosphate (RTP), which has at most 1/10 the absorption coefficient at
1064 nm of the lithium niobate (LiNbO$_3$) crystal from Initial
LIGO. At 200~W the RTP should produce a thermal lens of 200 m and
higher order mode content of less than 1\%, compared to the 3.3~m lens
the LiNbO$_3$ produces at 10~W. The RTP has a minimal risk of damage,
because it has both twice the damage threshold of LiNbO$_3$ and is
subjected to a beam twice the size of that in Initial LIGO. RTP and
LiNbO$_3$ have similar electro-optic coefficients. Also, RTP's $dn/dT$
anisotropy is 50\% smaller. Table \ref{tab:EOMcrystals} compares the
properties of most interest of the two crystals.

\begin{table*}
\caption{Comparison of selected properties of the Initial and Enhanced
 LIGO EOM crystals, LiNbO$_3$ and RTP, respectively. RTP was
 preferred for Enhanced LIGO because of its lower absorption,
 superior thermal properties, and similar 
 electro-optic properties \citep{UFLIGOGroup2006Upgrading}.}  
\centering
\begin{tabular}{l l l l}
\hline\hline
 & units & LiNbO$_3$ & RTP \\
\hline
damage threshold & MW/cm$^2$ & 280 & $>600$ \\
absorption coeff. at 1064 nm & ppm/cm & $< 5000$ & $< 500$ \\
electro-optic coeff. ($n_z^3 r_{33}$) & pm/V & 306 & 239 \\
$dn_y/dT$ & 10$^{-6}$/K & 5.4 & 2.79 \\
$dn_z/dT$ & 10$^{-6}$/K & 37.9 & 9.24 \\
\hline\hline
\end{tabular}
\label{tab:EOMcrystals}
\end{table*}

\begin{figure}
\begin{centering}
\includegraphics{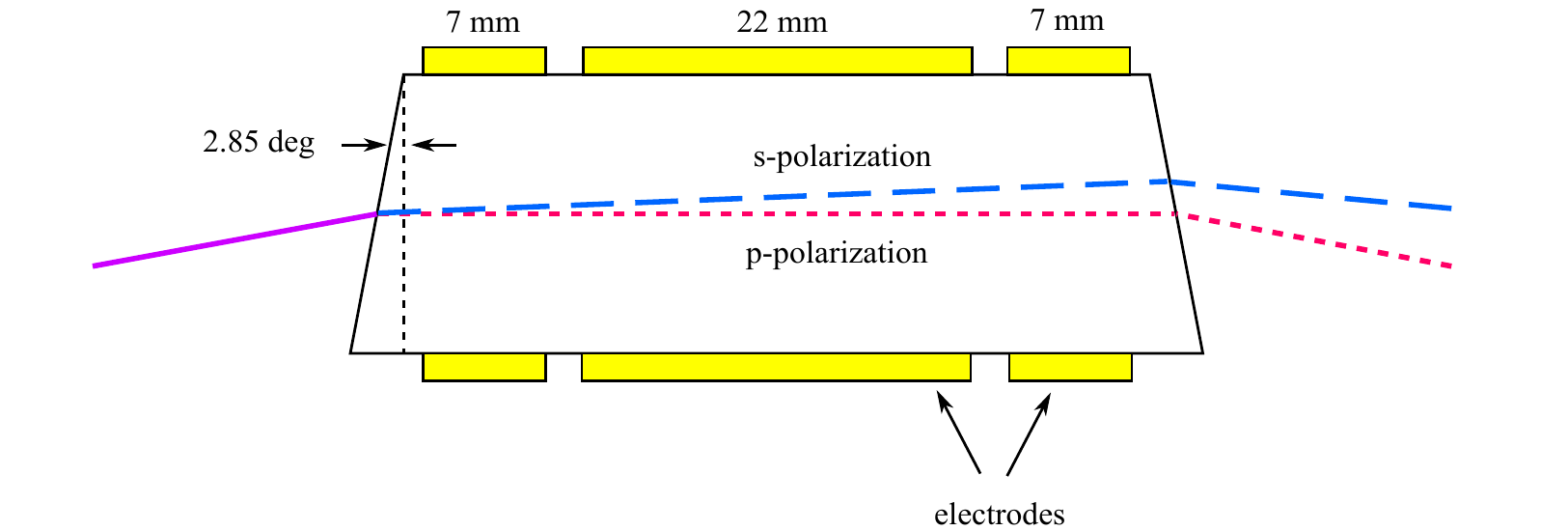}
\includegraphics{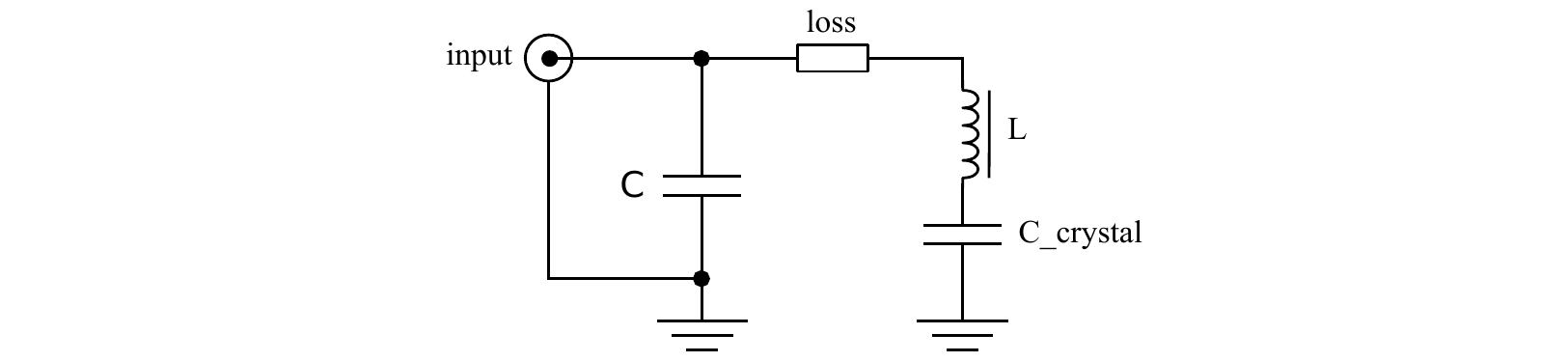}
%\subfigure{\includegraphics{figures/EOMthesis.pdf}}
%\subfigure{\includegraphics{figures/EOMcircuit_thesis.pdf}}
\caption{(Color online) Electro-optic modulator design. (a) The single
  RTP crystal is sandwiched between three sets of electrodes that
  apply three different modulation frequencies. The wedged ends of the
  crystal separate the polarizations of the light. The p-polarized
  light is used in the interferometer. (b) A schematic for each of the
  three impedance matching circuits of the EOM. For the three sets of
  electrodes, each of which creates its own $C_{crystal}$, a capacitor
  is placed parallel to the LC circuit formed by the crystal and a
  hand-wound inductor.  The circuits provide 50~$\Omega$ input
  impedance on resonance and are housed in a separate box from the
  crystal.}
\label{fig:EOM}
\end{centering}
\end{figure}

We procured the RTP crystals from Raicol and packaged them into
specially-designed, custom-built modulators. The crystal dimensions are
$4 \times 4 \times 40$ mm and their faces are wedged by $2.85^\circ$
and anti-reflection (AR) coated. The wedge serves to separate the
polarizations and prevents an etalon effect, resulting in a
suppression of amplitude modulation. Only one crystal is used in the
EOM in order to reduce the number of surface reflections. Three
separate pairs of electrodes, each with its own resonant LC circuit,
are placed across the crystal in series, producing the three required
sets of RF sidebands: 24.5~MHz, 33.3~MHZ and 61.2~MHz. A diagram is
shown in Fig. \ref{fig:EOM}. Reference
\citep{Quetschke2008ElectroOptic} contains further details about the
modulator architecture.

\subsection{Mode cleaner design}
The MC is a suspended 12.2~m long triangular ring cavity
with finesse $\mathcal{F}$=1280 and free spectral range of
12.243~MHz. The three mirror architecture was selected over the
standard two mirror linear filter cavity because it acts as a
polarization filter and because it eliminates direct path back
propagation to the laser \citep{Raab1992Estimation}. A pick-off of the
reflected beam is naturally facilitated for use in generating control
signals. A potential downside to the three mirror design is the
introduction of astigmatism, but this effect is negligible due to the
small opening angle of the MC. 

The MC has a round-trip length of 24.5~m. The beam waist has a radius of
1.63~mm and is located between the two 45$^\circ$ flat mirrors, MC1 and
MC3. See Figure~\ref{fig:IOschematic}. A concave third mirror, MC2,
18.15~m in radius of curvature, forms the far point of the mode
cleaner's isosceles triangle shape. The power stored in the MC is 408
times the amount coupled in, equivalent to about 2.7~kW in Initial
LIGO and at most 11~kW for Enhanced LIGO. The peak irradiances are
32~kW/cm$^2$ and 132~kW/cm$^2$ for Initial LIGO and Enhanced LIGO,
respectively. 

The MC mirrors are 75 mm in diameter and 25 mm thick. The
substrate material is fused silica and the mirror coating is made of
alternating layers of silica and tantala. In order to reduce the
absorption of light in these materials and therefore improve the
transmission and modal quality of the beam in the MC, we
removed particulate by drag wiping the surface of the mirrors with
methanol and optical tissues. The MC was otherwise identical
to that in Initial LIGO.

\subsection{Faraday isolator design}
The Enhanced LIGO FI design required not only the use of
low absorption optics, but additional design choices to mitigate any
residual thermal lensing and birefringence. In addition, trade-offs
between optical efficiency in the forward direction, optical isolation
in the backwards direction, and feasibility of physical access of the
return beam for signal use were considered. The result is that the
Enhanced LIGO FI needed a completely new architecture
and new optics compared to both the Initial LIGO FI and commercially
available isolators.

\begin{figure}
\begin{centering}
\includegraphics[width=0.9\columnwidth]{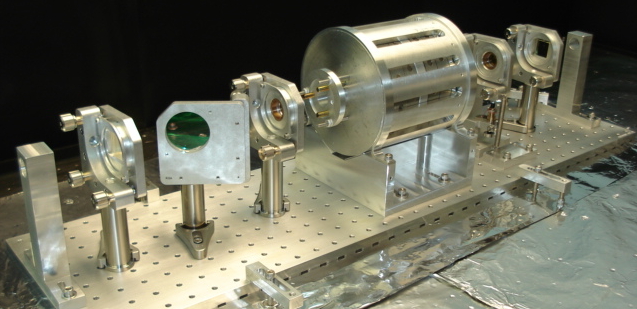}
\includegraphics{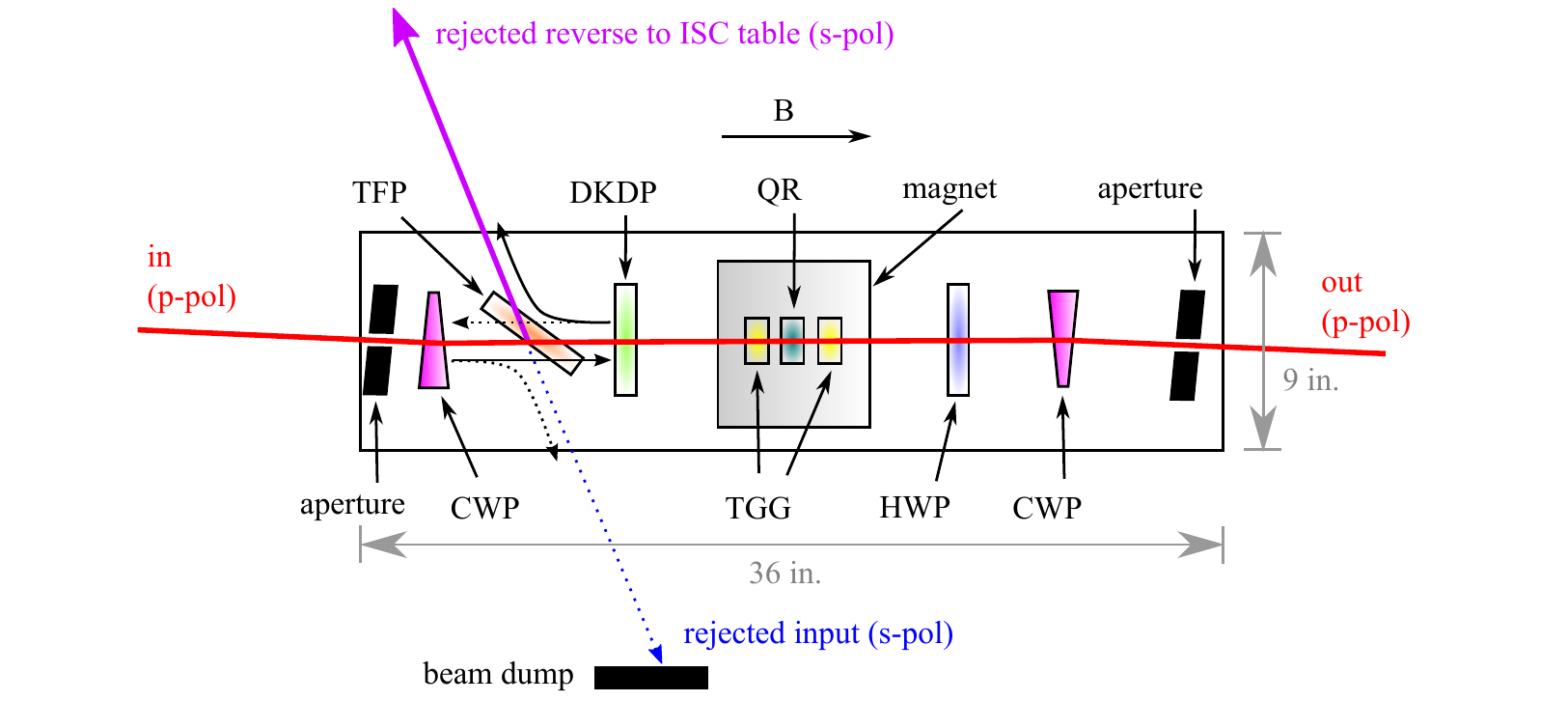}
\caption{(Color online) Faraday isolator photograph and schematic. The
  FI preserves the polarization of the light in the
  forward-going direction and rotates it by 90 degrees in the reverse
  direction. Light from the MC enters from the left and exits at the
  right towards the interferometer. It is ideally p-polarized, but any
  s-polarization contamination is promptly diverted $\sim 10$ mrad by
  the CWP and then reflected by the TFP and dumped. The p-polarized
  reflected beam from the interferometer enters from the right and is
  rotated to s-polarized light which is picked-off by the TFP and sent
  to the Interferometer Sensing and Control (ISC) table. Any
  imperfections in the Faraday rotation of the interferometer return
  beam results in p-polarized light traveling backwards along the
  original input path.}
\label{fig:FI}
\end{centering}
\end{figure}

Figure \ref{fig:FI} shows a photograph and a schematic of the Enhanced
LIGO FI. It begins and ends with low absorption calcite
wedge polarizers (CWP). Between the CWPs is a thin film polarizer
(TFP), a deuterated potassium dihydrogen phosphate (DKDP) element, a
half-wave plate (HWP), and a Faraday rotator. The rotator is made of
two low absorption terbium gallium garnet (TGG) crystals sandwiching a
quartz rotator (QR) inside a 7-disc magnet with a maximum field
strength of 1.16~T. The forward propagating beam upon passing through
the TGG, QR, TGG, and HWP elements is rotated by $+22.5^\circ -
67.5^\circ + 22.5^\circ + 22.5^\circ = 0^\circ$. In the reverse
direction, the rotation through HWP, TGG, QR, TGG is $-22.5^\circ +
22.5^\circ + 67.5^\circ + 22.5^\circ = 90^\circ$. The TGG crystals are
non-reciprocal devices while the QR and HWP are reciprocal.

\subsubsection{Thermal birefringence} 
Thermal birefringence is addressed in the Faraday rotator by the use
of the two TGG crystals and one quartz rotator rather than the typical
single TGG \citep{Khazanov2000Suppression}.  In this configuration,
any thermal polarization distortions that the beam experiences while
passing through the first TGG rotator will be mostly undone upon
passing through the second. The multiple elements in the magnet
required a larger magnetic field than in Initial LIGO. The 7-disc
magnet is 130~mm in diameter and 132~mm long and placed in housing
155~mm in diameter and 161~mm long. The TGG diameter is 20~mm.

\subsubsection{Thermal lensing}  
Thermal lensing in the FI is addressed by including
DKDP, a negative $dn/dT$ material, in the beam path. Absorption of
light in the DKDP results in a de-focusing of the beam, which
partially compensates for the thermal focusing induced by absorption
in the TGGs \citep{Mueller2002Method, Khazanov2004Compensation}.  The
optical path length (thickness) of the DKDP is chosen to slightly
over-compensate the positive thermal lens induced in the TGG crystals,
anticipating other positive thermal lenses in the system.

\subsubsection{Polarizers}  
The polarizers used (two CWPs and one TFP) each offer advantages and
disadvantages related to optical efficiency in the forward-propagating
direction, optical isolation in the reflected direction, and thermal
beam drift. The CWPs have very high extinction ratios ($>10^5$) and
high transmission ($>$ 99\%) contributing to good optical efficiency
and isolation performance. However, the angle separating the exiting
orthogonal polarizations of light is very small, on the order of 10
mrad. This small angle requires the light to travel relatively large
distances before we can pick off the beams needed for interferometer
sensing and control. In addition, thermally induced index of
refraction gradients due to the 4.95$^{\circ}$ wedge angle of the CWPs
result in thermal drift. However, the CWPs for the Enhanced LIGO FI
have a measured low absorption of 0.0013 cm$^{-1}$ with an expected
thermal lens of 60~m at 30~W and drift of less than 1.3
\microradnospace/W \citep{UFLIGOGroup2006Upgrading}.

The advantages of the thin film polarizer over the calcite wedge
polarizer are that it exhibits negligible thermal drift when compared
with CWPs and it operates at the Brewster angle of 55$^\circ$, thus
diverting the return beam in an easily accessible way. However, the
TFP has a lower transmission than the CWP, about 96\%, and an
extinction ratio of only 10$^3$.

Thus, the combination of CWPs and a TFP combines the best of each to
provide a high extinction ratio (from the CWPs) and ease of reflected
beam extraction (from the TFP). The downsides that remain when using
both polarizers are that there is still some thermal drift from the
CWPs. Also the transmission is reduced due to the TFP and to the fact
that there are 16 surfaces from which light can scatter.

\subsubsection{Heat conduction}
\label{sec:heatconduction}
Faraday isolators operating in a vacuum environment suffer from
increased heating with respect to those operating in air. Convective
cooling at the faces of the optical components is no longer an
effective heat removal channel, so proper heat sinking is essential to
minimize thermal lensing and depolarization. It has been shown that
Faraday isolators carefully aligned in air can experience a dramatic
reduction in isolation ratio ($>$ 10-15 dB) when placed in vacuum
\citep{TheVIRGOCollaboration2008Invacuum}. The dominant cause is the
coupling of the photoelastic effect to the temperature gradient
induced by laser beam absorption. Also of importance is the
temperature dependence of the Verdet constant--different spatial parts
of the beam experience different polarization rotations in the
presence of a temperature gradient \cite{Barnes1992Variation}.

To improve heat conduction away from the Faraday rotator optical
components, we designed a housing for the TGG and quartz crystals that
provided improved heat sinking to the Faraday rotator. We wrapped the
TGGs with indium foil that made improved contact with the housing and
we cushioned the DKDP and the HWP with indium wire in their aluminum
holders. This has the additional effect of avoiding the development of
thermal stresses in the crystals, an especially important
consideration for the very fragile DKDP.

\subsection{Mode-matching telescope design}
% from May 31, 2007 elog
The mode matching into the interferometer (at Livingston) was measured
to be at best 90\% in Initial LIGO. Because of the stringent
requirements placed on the LIGO vacuum system to reduce phase noise
through scattering by residual gas, standard opto-mechanical
translators are not permitted in the vacuum; it is therefore not
possible to physically move the mode matching telescope mirrors while
operating the interferometer. Through a combination of needing to move
the MMTs in order to fit the new FI on the in-vacuum optics table and
additional measurements and models to determine how to improve the
coupling, a new set of MMT positions was chosen for Enhanced
LIGO. Fundamental design considerations are discussed in
Ref. \citep{Delker1997Design}.

\section{Performance of the Enhanced LIGO Input Optics}
\label{sec:performance}
The most convincing figure of merit for the IO performance is that the
Enhanced LIGO interferometers achieved low-noise operation with 20 W
input power without thermal issues from the IO. Additionally, the IO
were operated successfully up to the available 30 W of power.
(Instabilities with other interferometer subsystems limited the
Enhanced LIGO science run operation to 20~W.)

We present in this section detailed measurements of the IO performance
during Enhanced LIGO. Specific measurements and results presented in
figures and the text come from Livingston; performance at Hanford was
similar and is included in tables summarizing the results.

\subsection{Optical efficiency}
The optical efficiency of the Enhanced LIGO IO from EOM to
recycling mirror was 75\%, a marked improvement over the approximate
60\% that was measured for Initial LIGO. A substantial part of the
improvement came from the discovery and subsequent correction of a
6.5\% loss at the second of the in-vacuum steering mirrors directing
light into the MC (refer to Fig. \ref{fig:IOschematic}). A 45$^\circ$
reflecting mirror had been used for a beam with an 8$^\circ$ angle of
incidence. Losses attributable to the MC and FI are described in the following sections. A summary of the IO
power budget is found in Table \ref{tab:pwrbudget}.

\begin{table*}
\caption{Enhanced LIGO IO power budget. Errors are $\pm
    1\%$, except for the TFP loss whose error is $\pm 0.1\%$. The
    composite MC transmission is the percentage of power after the MC to
    before the MC and is the product of the MC visibility and
    transmission. Initial LIGO values,
    where known, are included in parentheses and have errors of several
    percent.} 
\centering
\begin{tabular}{l l l}
\hline\hline
 & Livingston & Hanford \\
\hline
MC visibility & 92\% & 97\% \\
MC transmission & 88\% & 90\% \\
Composite MC transmission & 81\% (72\%) & 87\% \\
FI transmission &       93\% (86\%) & 94\% (86\%) \\
\hspace{0.5cm} - TFP loss & 4.0\% & 2.7\% \\ 
IO efficiency (PSL to RM) & 75\% (60\%) & 82\% \\
\hline\hline
\end{tabular}
\label{tab:pwrbudget}
\end{table*}

\subsubsection{Mode cleaner losses} 
The MC was the greatest single source of power loss in both
Initial and Enhanced LIGO. The MC visibility,
\begin{equation}
V = \frac{P_{\mathrm{in}} - P_{\mathrm{refl}}}{P_{\mathrm{in}}},
\label{eq:vis}
\end{equation}
where $P_{\mathrm{in}}$ is the power injected into the MC and
$P_{\mathrm{refl}}$ the power reflected, was 92\%. Visibility
reduction is the result of higher order mode content of
$P_{\mathrm{in}}$ and mode mismatch into the MC. The visibility was
constant within 0.04\% up to 30~W input power at both sites, providing
a positive indication that thermal aberrations in the MC and upstream
were negligible.

88\% of the light coupled into the MC was transmitted. 2.6\% of these
losses were caused by poor AR coatings on the second surfaces of the
$45^\circ$ MC mirrors. The measured surface microroughness of
$\sigma_{rms}< 0.4$ nm \cite{1998Component} caused scatter losses of 
$[4 \pi \sigma_{rms}/\lambda]^2 < 22$ ppm per mirror inside the MC, or
a total of 2.7\% losses in transmission. 

Another source of MC losses is via absorption of heat by particulates
residing on the mirror's surface. We measured the absorption with a
technique that makes use of the frequency shift of the thermally
driven drumhead eigenfrequencies of the mirror substrate
\citep{Punturo2007Mirror}. The frequency shift directly correlates
with the MC absorption via the substrate's change in Young's modulus
with temperature, $dY/dT$. A finite element model (COMSOL
\cite{COMSOL}) was used to compute the expected frequency shift from a
temperature change of the substrate resulting from the mirror coating
absorption. The measured eigenfrequencies for each mirror at room
temperature are 28164~Hz, 28209~Hz, and 28237~Hz, respectively.

\begin{figure}
\begin{centering}
\includegraphics[width=1.0\columnwidth]{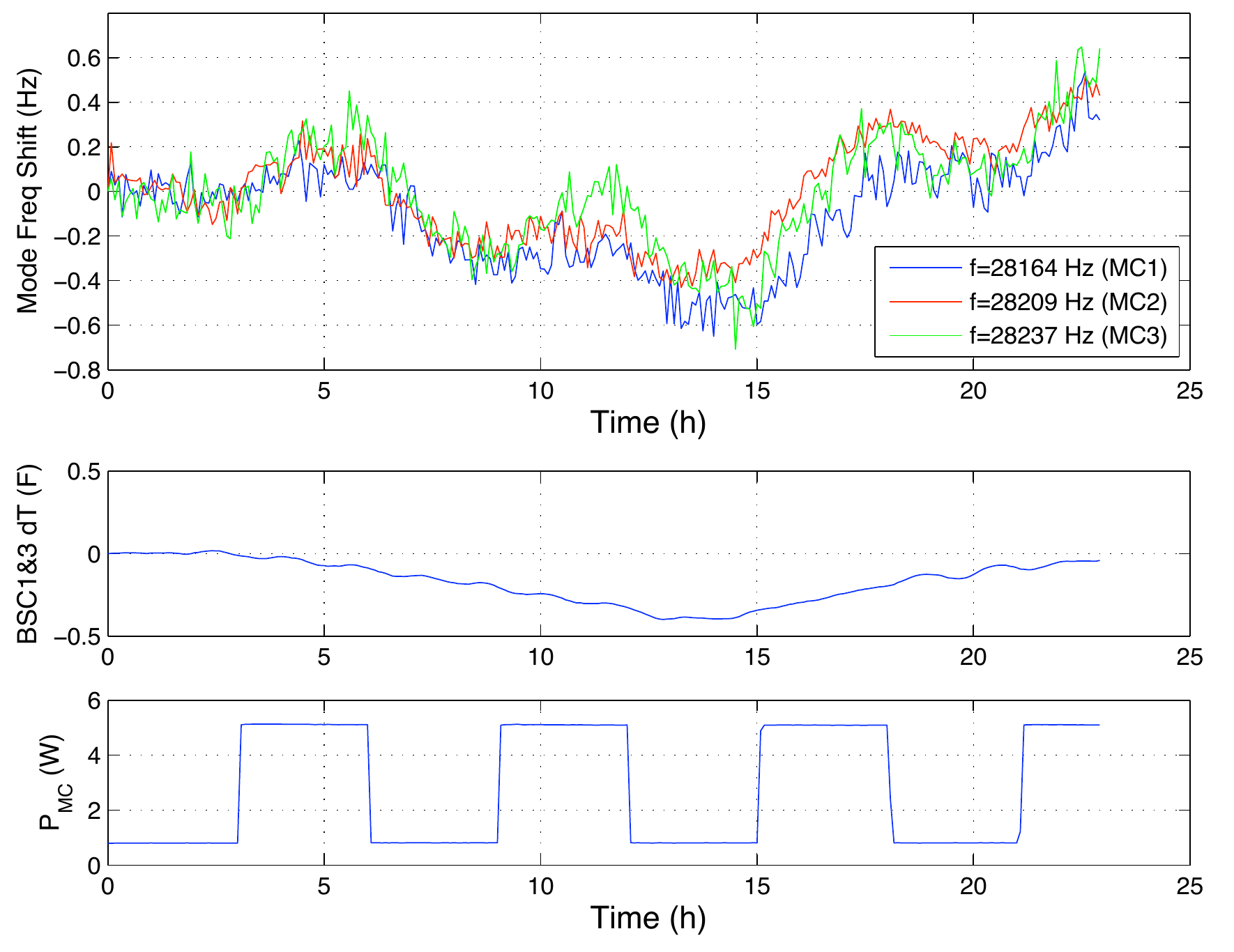}
\caption{(Color online) Data from the MC absorption measurement. Power into
the MC was cycled between 0.9~W and 5.1~W at 3 hour intervals (bottom
frame) and the change in frequency of the drumhead mode of each mirror
was recorded (top frame). The ambient temperature (middle frame) was
also recorded in order to correct for its effects.} 
\label{fig:MCabsorption}
\end{centering}
\end{figure}

We cycled the power into the MC between 0.9~W and 5.1~W at 3
hour intervals, allowing enough time for a thermal characteristic time
constant to be reached.  At the same time, we recorded the
frequencies of the high Q drumhead mode peaks as found in the mode
cleaner frequency error signal, heterodyned down by 28~kHz. See Figure
\ref{fig:MCabsorption}. Correcting for ambient temperature
fluctuations, we find a frequency shift of 0.043, 0.043, and 0.072
Hz/W. As a result of drag-wiping the mirrors, the absorption decreased
for all but one mirror, as shown for both Hanford and Livingston in
Table~\ref{tab:MCabsorption2}.

\begin{table}
\caption{Absorption values for the Livingston and Hanford mode
 cleaner mirrors before (in parentheses) and after drag wiping. The precision is $\pm 10\%$.} 
\centering
\begin{tabular}{l l l}
\hline\hline
mirror & Livingston & Hanford\\
\hline
MC1 & 2.1 ppm (18.7 ppm) & 5.8 (6.1 ppm) \\
MC2 & 2.0 ppm (5.5 ppm) & 7.6 (23.9 ppm) \\
MC3 & 3.4 ppm (12.8 ppm) & 15.6 (12.5 ppm) \\
\hline\hline
\end{tabular}
\label{tab:MCabsorption2}
\end{table}

\subsubsection{Faraday isolator losses} 
The FI was the second greatest source of power loss with
its transmission of 93\%. This was an improvement over the
86\% transmission of the Initial LIGO FI. The most lossy element in the
FI is the thin film polarizer, accounting for 4\% of
total losses. The integrated losses from AR coatings and absorption in the
TGGs, CWPs, HWP, and DKDP account for the remaining 3\% of missing power.

\subsection{Faraday isolation ratio}
The isolation ratio is defined as the ratio of power incident on the
FI in the reverse direction (the light reflected from the
interferometer) to the power transmitted in the reverse direction and
is often quoted in decibels: isolation ratio~=~$10
\log_{10}(P_{\mathrm{in-reverse}}/P_{\mathrm{out-reverse}})$.  We
measured the isolation ratio of the FI as a function of
input power both in air prior to installation and \emph{in situ}
during Enhanced LIGO operation.

To measure the in-vacuum isolation ratio, we misaligned the
interferometer arms so that the input beam would be promptly reflected
off of the $97\%$ reflective recycling mirror. This also has the
consequence that the FI is subjected to twice the input
power. Our isolation monitor was a pick-off of the backwards
transmitted beam taken immediately after transmission through the
FI that we sent out of a vacuum chamber viewport. Refer to the
``isolation check beam'' in Fig.~\ref{fig:IOschematic}. The in air
measurement was done similarly, except in an optics lab with a
reflecting mirror placed directly after the FI.

\begin{figure}
\begin{centering}
\includegraphics[width=1.0\columnwidth]{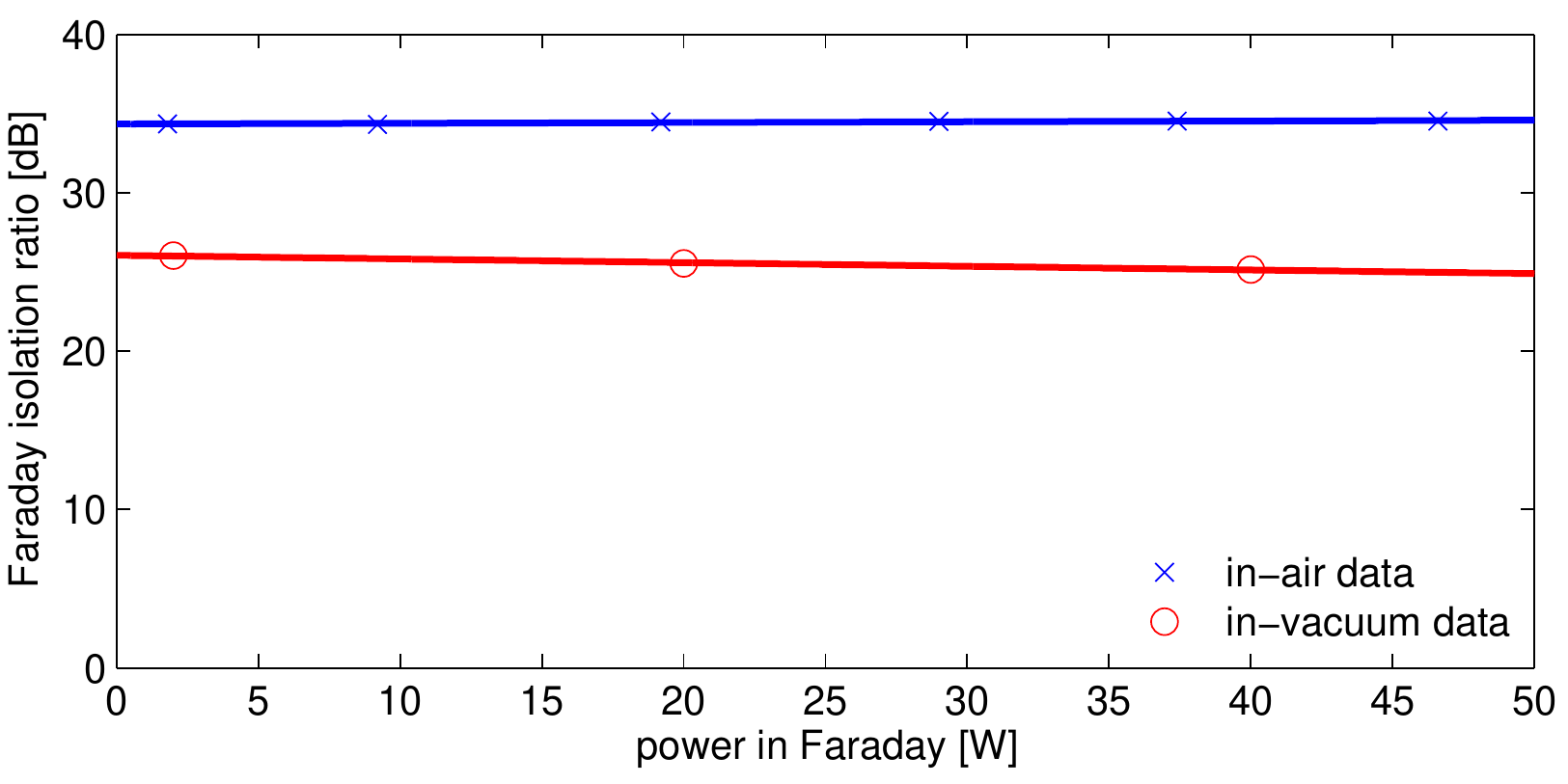}
\caption{Faraday isolator isolation ratio as measured in air prior to
  installation and \emph{in situ} in vacuum. The isolation worsens by
  a factor of 6 upon placement of the FI in vacuum. The linear
  fits to the data show a constant in-air isolation ratio and an
  in-vacuum isolation ratio degradation of 0.02 dB/W.}
\label{fig:IR}
\end{centering}
\end{figure}

Figure \ref{fig:IR} shows our isolation ratio data. Most notably, we
observe an isolation decrease of a factor of six upon placing the
FI in vacuum, a result consistent with that reported by
Ref. \citep{TheVIRGOCollaboration2008Invacuum}. In air the isolation
ratio is a constant 34.46 $\pm$ 0.04 dB from low power up to 47~W, and
in vacuum the isolation ratio is 26.5 dB at low power. The underlying
cause is the absence of cooling by air convection. If we attribute the
loss to the TGGs, then based on the change in TGG polarization
rotation angle necessary to produce the measured isolation drop of
8~dB and the temperature dependence of the TGG's Verdet constant, we
can put an upper limit of 11~K on the crystal temperature rise from
air to vacuum. Furthermore, a degradation of 0.02~dB/W is measured in
vacuum.

\subsection{Thermal steering}
We measured the \emph{in situ} thermal angular drift of both the beam
transmitted through the MC and of the reflected beam from
the FI with up to 25~W input power. Just as for the
isolation ratio measurement, we misaligned the interferometer arms so
that the input beam would be promptly reflected off of the recycling
mirror. The Faraday rotator was thus subjected to up to 50~W total
and the MC to 25~W. 

Pitch and yaw motion of the MC transmitted and
interferometer reflected beams were recorded using the quadrant
photodiode (QPD) on the IO table and the RF alignment
detectors on the Interferometer Sensing and Control table (see
Fig. \ref{fig:IOschematic}). There are no lenses between the MC waist
and its measurement QPD, so only the path length between the two were
needed to calibrate in radians the pitch and yaw signals on the
QPD. The interferometer reflected beam, however, passes through
several lenses. Thus, ray transfer matrices and the two alignment
detectors were necessary to determine the Faraday drift calibration.

\begin{figure}
\begin{centering}
\includegraphics[width=1.0\columnwidth]{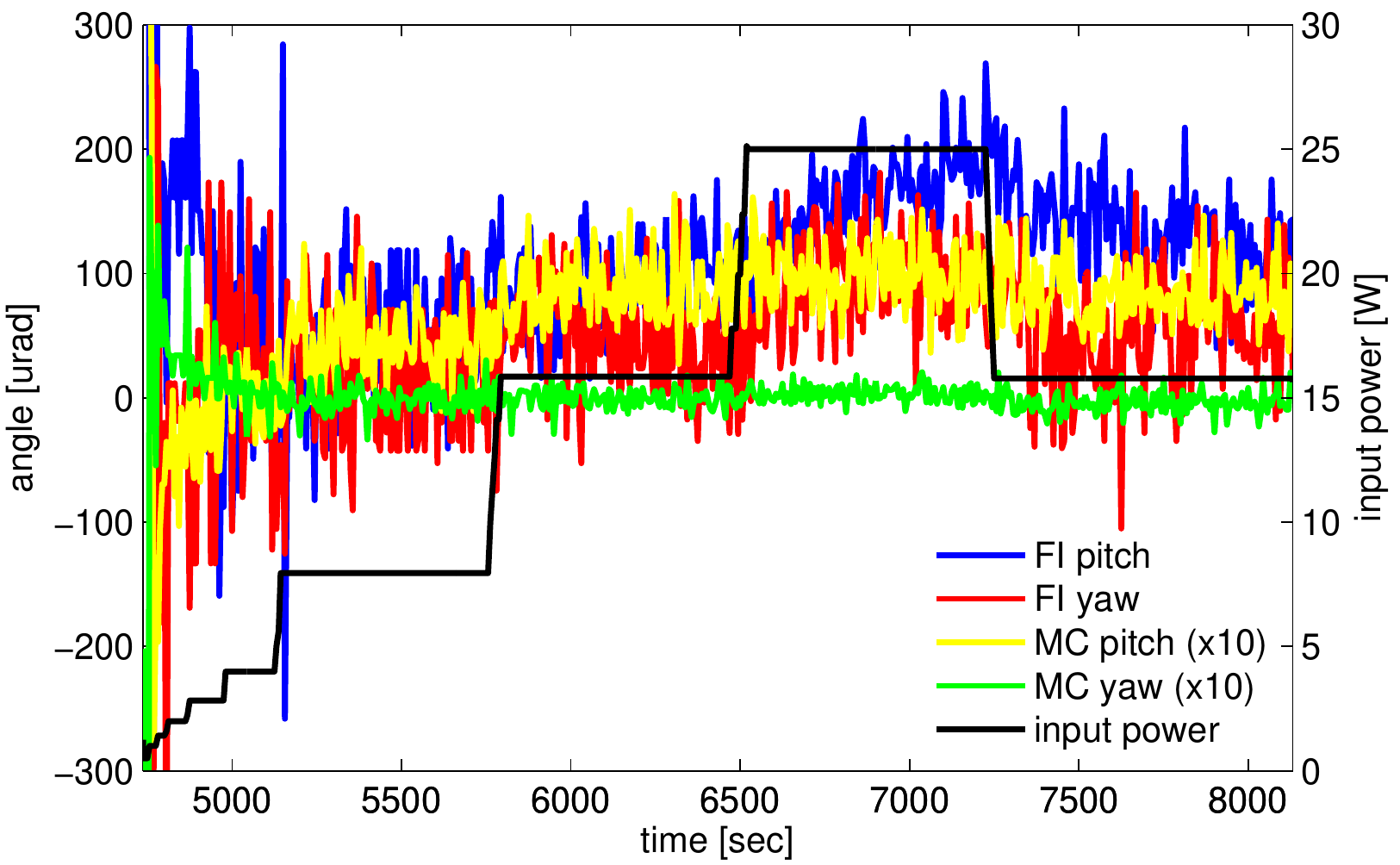}
\includegraphics[width=1.0\columnwidth]{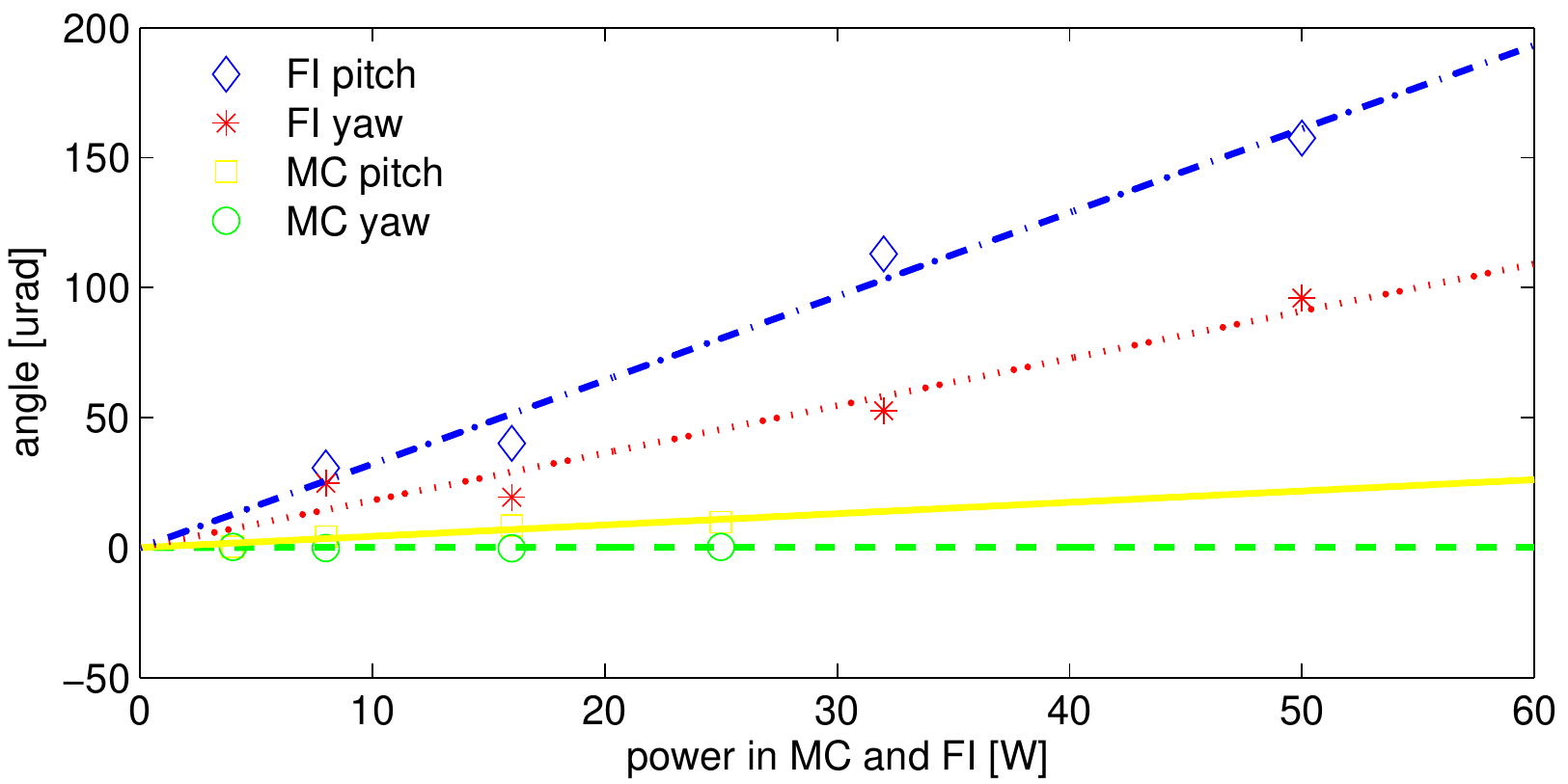}
\caption{(Color online) Mode cleaner and Faraday isolator thermal
  drift data. (a) Angular motion of the beam at the MC waist and FI
  rotator as the input power is stepped. The beam is double-passed
  through the Faraday isolator, so it experiences twice the input
  power. (b) Average beam angle per power level in the MC and
  FI. Linear fits to the data are also shown. The slopes for MC yaw, MC pitch, FI yaw,
  and FI pitch, respectively, are 0.0047, 0.44, 1.8, and 3.2 \microradnospace/W.} 
\label{fig:drift}
\end{centering}
\end{figure}

Figure \ref{fig:drift} shows the calibrated beam steering data. The
angle of the beam out of the MC does not change measurably
as a function of input power in yaw (4.7~nrad/W) and changes by only
440~nrad/W in pitch. For the FI, we record a beam drift
originating at the center of the Faraday rotator of 1.8~\microradnospace/W in
yaw and 3.2~\microradnospace/W in pitch. Therefore, when ramping the input
power up to 30~W during a full interferometer lock, the upper limit on
the drift experienced by the reflected beam is about 100
\microradnospace. This is a thirty-fold reduction with respect to the Initial
LIGO FI and represents a fifth of the beam's divergence
angle, $\theta_{div}$~=~490 \microradnospace.

\subsection{Thermal lensing}
We measured the profiles of both the beam transmitted through the mode
cleaner and the reflected beam picked off by the FI at low
($\sim$~1~W) and high ($\sim$~25~W) input powers to assess the degree
of thermal lensing induced in the MC and FI. Again, we misaligned the
interferometer arms so that the input beam would be promptly reflected
off the recycling mirror. We picked off a fraction of the reflected
beam on the Interferometer Sensing and Control table and of the mode
cleaner transmitted beam on the IO table (refer to
Fig. \ref{fig:IOschematic}), placed lenses in each of their paths, and
measured the beam diameters at several locations on either side of the
waists created by the lenses. A change in the beam waist size or
position as a function of laser power indicates the presence of a
thermal lens.

\begin{figure}
\begin{centering}
\includegraphics[width=1.0\columnwidth]{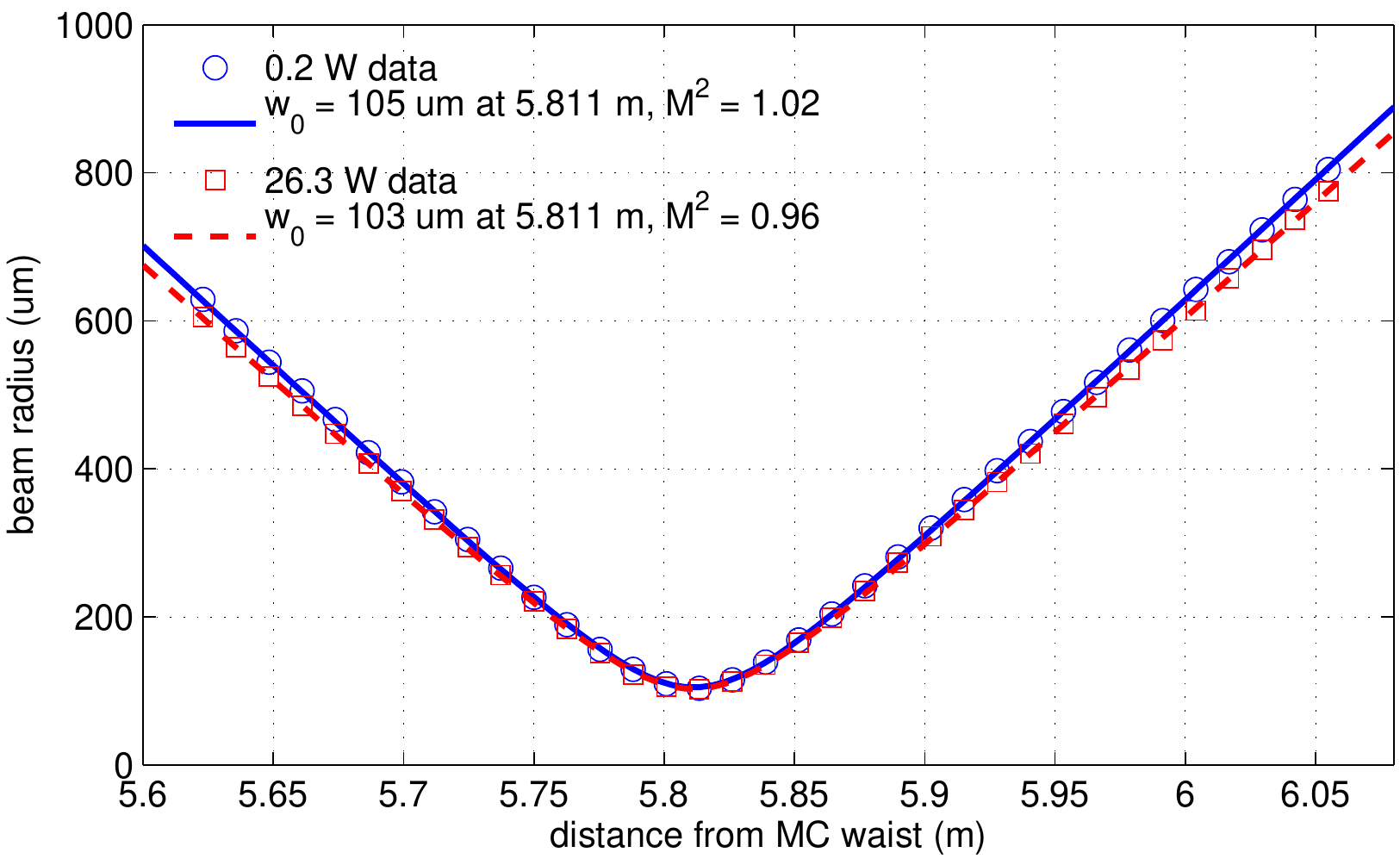}
\caption{(Color online) Profile at high and low powers of a pick-off of the beam
  transmitted through the MC. The precision of the beam
  profiler is $\pm 5\%$. Within the error of the measurement, there
  are no obvious degradations.} 
\label{fig:MC_lensing}
\end{centering}
\end{figure}

\begin{figure}
\begin{centering}
\includegraphics[width=1.0\columnwidth]{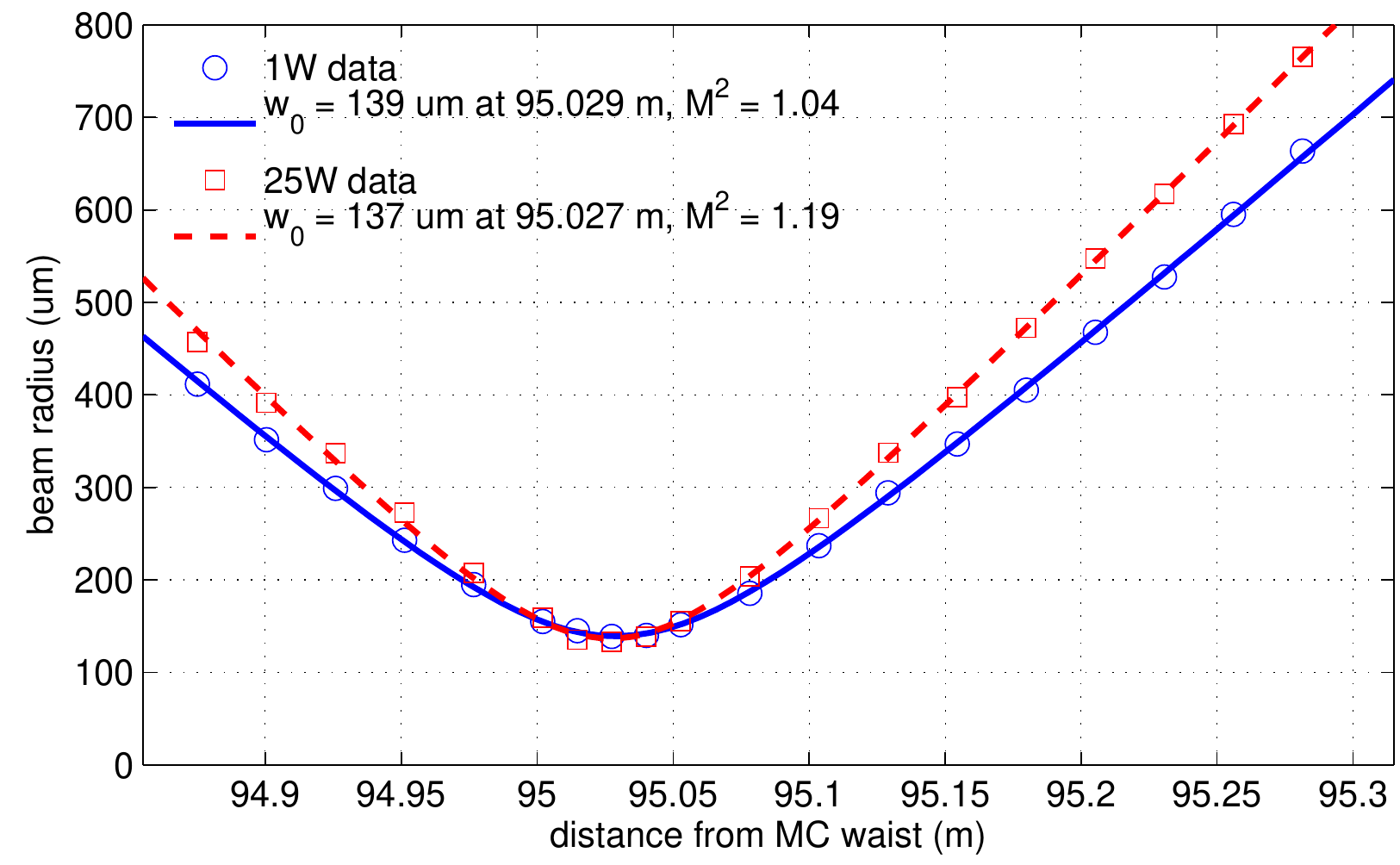}
\caption{(Color online) Faraday isolator thermal lensing data. With 25
  W into the Faraday isolator (corresponding to 50 W in double pass),
  the beam has a steeper divergence than a pure TEM$_{00}$ beam,
  indicating the presence of higher order modes. Errors are $\pm
  5.0\%$ for each data point.}
\label{fig:FI_lensing}
\end{centering}
\end{figure}

As seen in Fig. \ref{fig:MC_lensing} and \ref{fig:FI_lensing}, the
waists of the two sets of data are collocated: no thermal lens is
measured. For the FI, the divergence of the low and high
power beams differs, indicating that the beam quality degrades with
power. The $M^2$ factor at 1~W is 1.04 indicating the beam is nearly
perfectly a TEM$_{00}$ mode. At 25~W, $M^2$ increases to 1.19,
corresponding to increased higher-order-mode content. The percentage
of power in higher-order modes depends strongly on the mode order and
relative phases of the modes, and thus cannot be determined from this
measurement \citep{Kwee2007Laser}.

The results for the MC are consistent with no thermal
lensing. The high and low power beam profiles are within each other's
error bars and well below our requirements.

\begin{figure}
\begin{centering}
\includegraphics[width=1.0\columnwidth]{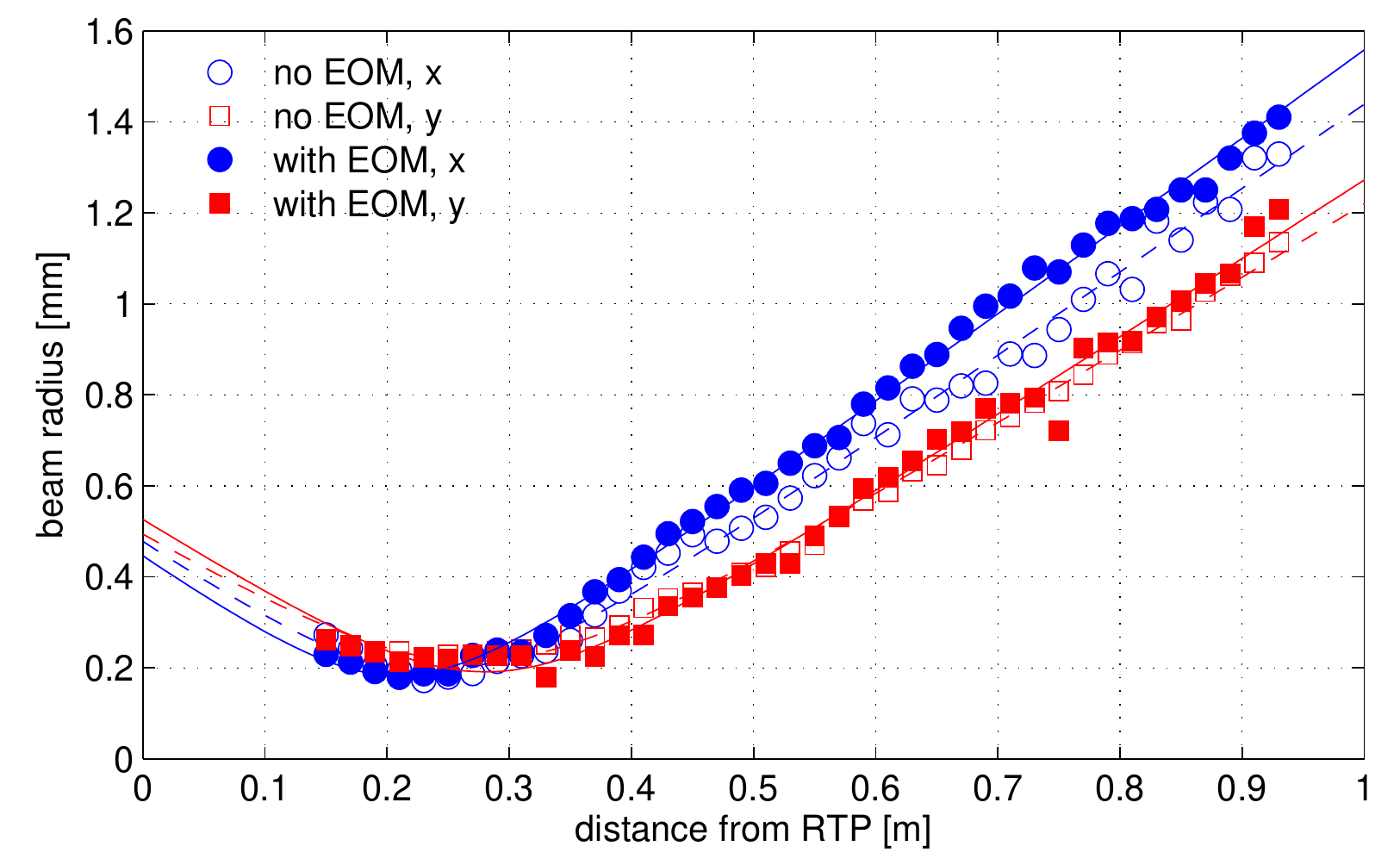}
\caption{(Color online) EOM thermal lensing data. The x- and
  y-direction beam profiles with 160~W through the EOM (closed circles
  and squares) place a lower limit of 4~m on the induced thermal lens
  when compared to the beam profiles without the EOM (open circles and
  squares).}
\label{fig:EOMlensing}
\end{centering}
\end{figure}

We also measured the thermal lensing of the EOM prior to its
installation in Enhanced LIGO by comparing beam profiles of a 160~W
beam with and without the EOM in its path. The data for both
cross-sections of the beam is presented in
Fig.~\ref{fig:EOMlensing}. We observe no significant thermal lensing
in the y-direction and a small effect in the x-direction. An upper
limit for the thermal lens in the x-direction can be calculated to be
greater than 4~m, which is 10 times larger than the Rayleigh range of
the spatial mode. The mode matching degradation is therefore less than
1\%. Although a direct test for Advanced LIGO because of the power
used, this measurement also serves to demonstrate the effectiveness of
the EOM design for Enhanced LIGO powers.

\subsection{Mode-matching}
We measured the total interferometer visibility (refer to
Eq.~\ref{eq:vis}) as an indirect way of determining the carrier
mode-matching to the interferometer. In this case, $P_{\mathrm{in}}$
is the power in the reflected beam when the interferometer cavities
are unlocked and $P_{\mathrm{refl}}$ is the power in the reflected
beam when all of the interferometer cavities are on resonance.

The primary mechanisms that serve to reduce the interferometer
visibility from unity are: carrier mode-matching, carrier impedance
matching, and sideband light. We measured the impedance matching at
LLO to be $>$~99.5\%; impedance matching therefore makes a negligible
contribution to the power in the reflected beam. We also measured that
due to the sidebands, the carrier makes up 86\% of the power in the
reflected beam with the interferometer unlocked and 78\% with the
interferometer locked; to compensate, we reduce the total
$P_{\mathrm{refl}}/P_{\mathrm{in}}$ ratio by 10\%. With the
interferometer unlocked, there is also a 2.7\% correction for the
transmission of the RM.

Initially, anywhere between 10\% and 17\% of the light was rejected by
the interferometer due to poor, power-dependent mode matching.  After
translating the mode-matching telescope mirrors during a vacuum
chamber incursion and upgrading the other IO components, the mode
mismatch we measured was 8\% and independent of input power. The MMT
thus succeeds in coupling 92\% of the light into the interferometer at
all times, marking both an improvement in MMT mirror placement and
success in eliminating measurable thermal issues.

\section{Implications for Advanced LIGO}
\label{sec:aLIGO}
As with other Advanced LIGO interferometer components, Enhanced LIGO
served as a technology demonstrator for the Advanced LIGO Input
Optics, albeit at lower laser powers than will be used there. The
performance of the Enhanced LIGO IO components at 30~W of
input power allows us to infer their performance in Advanced LIGO.
The requirements for the Advanced LIGO IO demand are for
similar performance to Enhanced LIGO, but with almost 8 times the
laser power.

The Enhanced LIGO EOM showed no thermal lensing, degraded
transmission, nor damage in over 17,000 hours of sustained operation
at 30~W of laser power. Measurements of the thermal lensing in RTP at
powers up to 160 W show a relative power loss of $< 0.4\%$, indicating
that thermal lensing should be negligible in Advanced LIGO.  Peak
irradiances in the EOM will be approximately four times that of
Enhanced LIGO (a 45\% larger beam diameter will somewhat offset the
increased power).  Testing of RTP at 10 times the expected Advanced
LIGO irradiance over 100~hours show no signs of damage or degraded
transmission.

The MC showed no measurable change in operational state as a
function of input power.  This bodes well for the Advanced LIGO mode
cleaner.  Compared with the Enhanced LIGO MC, the Advanced
LIGO MC is designed with a lower finesse (520) than Initial
LIGO (1280).  For 150~W input power, the Advanced LIGO MC
will operate with 3 times greater stored power than Initial LIGO.  The
corresponding peak irradiance is 400~kW/m$^2$, well below the
continuous-wave coating damage threshold.  Absorption in the Advanced
LIGO MC mirror optical coatings has been measured at
0.5~ppm, roughly four times less than the best mirror coating
absorption in Enhanced LIGO, so the expected thermal loading due to
coating absorption should be reduced in Advanced LIGO.  The larger
Advanced LIGO MC mirror substrates and higher input powers
result in a significantly higher contribution to bulk absorption,
roughly 20 times Enhanced LIGO, however the expected thermal lensing
leads to small change ($< 0.5 \%$) in the output mode
\citep{Arain2007Note}.

The Enhanced LIGO data obtained from the FI allows us to make several
predictions about how it will perform in Advanced LIGO.  The measured
isolation ratio decrease of 0.02~dB/W will result in a loss of 3~dB
for a 150~W power level expected for Advanced LIGO relative to its
cold state.  However, the Advanced LIGO FI will employ an \emph{in
  situ} adjustable half wave plate which will allow for a partial
restoration of the isolation ratio. In addition, a new FI scheme to
better compensate for thermal depolarization and thus yield higher
isolation ratios will be implemented
\cite{Snetkov2011Compensation}. The maximum thermally induced angular
steering expected is 480 \microrad (using a drift rate of 3.2
\microradnospace/W), approximately equal to the beam divergence
angle. This has some implications for the Advanced LIGO length and
alignment sensing and control system, as the reflected FI beam is used
as a sensing beam. Operation of Advanced LIGO at high powers will
likely require the use of a beam stabilization servo to lock the
position of the reflected beam on the sensing photodiodes.  Although
no measurable thermal lensing was observed (no change in the beam
waist size or position), the measured presence of higher order modes
in the FI at high powers is suggestive of imperfect thermal lens
compensation by the DKDP.  This fault potentially can be reduced by a
careful selection of the thickness of the DKDP to better match the
absorbed power in the TGG crystals.

\section{Summary}
\label{sec:summary}
In summary, we have presented a comprehensive investigation of the
Enhanced LIGO IO, including the function, design, and
performance of the IO.  Several improvements to the design and
implementation of the Enhanced LIGO IO over the Initial LIGO IO have
lead to improved optical efficiency and coupling to the main
interferometer through a substantial reduction in thermo-optical
effects in the major IO optical components, including the
electro-optic modulators, mode cleaner, and Faraday isolator.  The IO
performance in Enhanced LIGO enables us to infer its performance in
Advanced LIGO, and indicates that high power interferometry will be
possible without severe thermal effects.

\begin{acknowledgments} 
  The authors thank R.~Adhikari for his wisdom and guidance, B.~Bland
  for providing lessons to K.~Dooley and D.~Hoak on how to handle the
  small optics suspensions, K.~Kawabe and N.~Smith-Lefebvre for their
  support at LHO, T.~Fricke for engaging in helpful discussions, and
  V.~Zelenogorsky and D.~Zheleznov for their assistance in preparing
  for the Enhanced LIGO IO installation. Additionally, the authors
  thank the LIGO Scientific Collaboration for access to the data. This
  work was supported by the National Science Foundation through grants
  PHY-0855313 and PHY-0555453. LIGO was constructed by the California
  Institute of Technology and Massachusetts Institute of Technology
  with funding from the National Science Foundation and operates under
  cooperative agreement PHY-0757058. This paper has LIGO Document
  Number LIGO-P1100056.
\end{acknowledgments}

\end{document}